\documentclass[onecolumn,aps,pre,preprint]{revtex4-2}

\usepackage{amssymb}
\usepackage{amsmath}
\usepackage{graphicx}
\usepackage{xcolor}
\begin{document}
\title{Casimir Interaction between Polydisperse Colloids Trapped at a Fluid Interface}
\author{Seyed Emad Mousavi}
\author{Ehsan Noruzifar}
\email{noruzifar@kntu.ac.ir}
\affiliation{Faculty of Physics, K.N. Toosi University of 
	Technology, P.O. Box: 15875-4416, Tehran, Iran}

\begin{abstract}
We investigate the effect of polydispersity on fluctuation-induced interactions between colloids trapped at a fluid interface. Using the scattering-matrix formalism, we 
calculate the interaction energy in two- and three-body systems under three mechanical boundary conditions. We find that size asymmetry can either suppress or 
enhance the many-body interaction compared to a monodisperse system, with the outcome depending on colloids mobility and separation. For fixed colloids, the 
interaction is suppressed at short separations but enhanced at large separations. In contrast, for mobile colloids, the interaction is predominantly enhanced at large 
distances but exhibits a competitive behavior in the near-field. This culminates in a large, multi-order-of-magnitude amplified sensitivity {to size asymmetry} 
for bobbing and tilting colloids at long range, highlighting a complex interplay between geometry and colloids mobility with significant implications for self-assembly.
\end{abstract}

\maketitle
	\section{Introduction}
	\label{sec:introduction}
	
	Colloidal particles trapped at fluid interfaces experience interactions that are absent in 
	bulk~\cite{binks2006colloidal, stamou2000long, vella2005cheerios, oettel2009effective}. These types of interactions 
	are caused by interface deformations, thermal fluctuations of the interface, confinement effects, and degrees of 
	freedom of colloidal particles~\cite{oettel2007importance,lehle2006effective, dominguez2005capillary, 
		dominguez_2007, dominguez_2018,danov2010capillary, zeng2012colloidal}. Such forces influence two-dimensional 
		pattern formation~\cite{colinet_2010_pattern, gosh2007}  and self-assembly~\cite{lin2003nanoparticle, 
			guzman2022} at interfaces.
	
	Fluctuation-induced interactions, also known as Casimir-like forces, affect particles at interfaces in a unique way. 
	These forces which arise from thermal capillary wave fluctuations, are significant because they are not pairwise 
	additive and also depend on geometry and boundary 
	conditions~\cite{golestanian1999fluctuation,nfr2009anis,nfr2013three,nfr2013sct}. The critical role of boundary 
	conditions in tuning fluctuation-induced forces has also been extensively studied in the context of critical Casimir 
	interactions~\cite{nellen2009tunability, gambassi2024critical, vasilyev2011critical}. Therefore, an interesting 
	aspect of Casimir interactions in such systems is the influence of polydispersity, i.e., differences in colloid size 
	or shape.
	
	Casimir-like interactions between colloids at interfaces can be studied by theoretical approaches such as 
	perturbative methods, effective field theory~\cite{deserno2012,yolcu2014,deserno2014}, or the scattering 
	formalism~\cite{nfr2013sct,nfr2013three}. The latter uses the scattering method to study the impact of colloids 
	shape and fluctuations on such interactions. This is based on the scattering formalism developed for quantum Casimir 
	forces~\cite{emig2009} for a wide range of geometries and boundary 
	conditions~\cite{emig2015,nfr2013,nfr2012,nfr2011}. 
	 
	Most theoretical and numerical studies assume systems with uniform particle 
	sizes. However, colloids can show a range of sizes in real-world physical 
	systems~\cite{liu2018,zhou_2020}. The importance of size effects has indeed 
	been highlighted in related systems, for instance, in the application of 
	critical Casimir forces for size-selective particle 
	purification~\cite{villanueva2021concentration}. In this work, we examine the 
	effect of size variation on the Casimir interaction between colloids trapped 
	at a fluid interface.
		
We apply the previously developed scattering formalism~\cite{nfr2013sct, nfr2013three} to systematically investigate the effect of size variation on the Casimir 
interaction for both two- and three-body configurations. Our aim is to utilize this established framework to explore a new and physically rich parameter space---
namely, the role of polydispersity under various mechanical boundary conditions. We explore three distinct scenarios: (a) fixed colloids, (b) bobbing-only 
	colloids, and (c) bobbing and tilting colloids. 
We demonstrate that particle mobility acts as a qualitative {switch}, fundamentally altering the system's response to size asymmetry. In the following sections, we will analyze the complex, separation-dependent behavior for each scenario and uncover the physical mechanisms---from near-field geometric effects to far-field modal purification---that govern this tunable interaction.	
	
	This work is organized as follows: In Sec.~\ref{sec:method}, we explain and review the scattering formalism that we use to calculate the Casimir interaction between colloids, and also we describe how introducing new degrees of freedom to the motion of colloids would change their scattering matrix. In Sec.~\ref{sec:numerical}, we describe the algorithm used to produce numerical results for two- and three-body systems. In this section, we also analyze the convergence behavior of the interaction energy. In Sec.~\ref{sec:discussion}, we present our numerical results for the interaction energy between two and three colloids with different sizes, and discuss how size 
	polydispersity impacts the 2-body Casimir interaction and also the non-pairwise part. 
	Finally, in Sec.~\ref{sec:summary}, we summarize our findings and discuss their implications for experimental studies and self-assembly in colloidal systems at fluid interfaces.

	\section{Method}
	\label{sec:method}
	
	To calculate the fluctuation-induced (Casimir-like) interactions between colloidal particles at fluid interfaces, we use the scattering formalism~\cite{nfr2013sct}. 
	This formalism enables quick calculation of Casimir energies for systems with known scattering matrices. It also easily handles various boundary conditions and colloidal dynamics. In our study, we model the colloidal particles as rigid Janus spheres, which are trapped at a fluid interface, such as between air and water, and are affected by capillary-wave fluctuations.
	
	At thermal equilibrium, the free energy of fluctuation-induced interactions is given by the determinant of a matrix. This matrix entails both the scattering properties of individual colloids and their interaction through the surrounding fluctuating field. The free energy expression reads~\cite{nfr2013sct}
	\begin{equation}
		\frac{\mathcal{F}}{k_\mathrm{B}T} = \frac{1}{2} \ln \det (\mathbf{1} - \mathbf{\widetilde{T}U})\,,
	\end{equation}
	where $\mathcal{F}$ is the interaction free energy, $k_\mathrm{B}$ is Boltzmann’s constant, $T$ is the temperature, $\mathbf{\widetilde{T}}$ is the block-diagonal T-matrix describing the scattering response of colloids to capillary-wave fluctuations, and $\mathbf{U}$ is the translation matrix, each entry in $\mathbf{U}$ describes how multipole fields centered on one particle are translated to another particle's frame of reference.
	
	\subsection*{Two-Body Casimir Energy}
	
	For a system of two colloids, the matrix structure simplifies, and the interaction free energy reduces to
	\begin{equation}
		\label{eq:F2}
		\frac{\mathcal{F}^{(2)}}{k_\mathrm{B}T} = \frac{1}{2} \ln \det 
		\begin{pmatrix}
			f_{11} & f_{12} \\
			f_{21} & f_{22}
		\end{pmatrix}\,,
	\end{equation}
	where each block $f_{ij}$ is defined by
	\begin{equation}
		\label{eq:f-elements}
		f_{ij} = \delta_{ij} \mathbf{1} + \widetilde{\mathbb{T}}^{i} \mathbb{U}^{ij}(1 - \delta_{ij})\,.
	\end{equation}
	Here, $\widetilde{\mathbb{T}}^{i}$ is the T-matrix of colloid $i$, and $\mathbb{U}^{ij}$ is the translation matrix that connects modes from colloid $i$ to colloid $j$.
	
	\subsection*{Three-Body Casimir Energy and Non-Additivity}
	
	In the case of three colloidal particles, the full Casimir interaction becomes more intricate due to the emergence of non-additive contributions. The total free energy is calculated by
	\begin{equation}
		\label{eq:F3}
		\frac{\mathcal{F}^{(3)}}{k_\mathrm{B}T} = \frac{1}{2} \ln \det 
		\begin{pmatrix}
			f_{11} & f_{12} & f_{13} \\
			f_{21} & f_{22} & f_{23} \\
			f_{31} & f_{32} & f_{33}
		\end{pmatrix}\,,
	\end{equation}
	with $f_{ij}$ again defined as in Eq.~(\ref{eq:f-elements}). Importantly, the Casimir interaction in this setting is {\it not pairwise additive} and the total energy cannot be written as a sum of interactions between particle pairs. Therefore, we isolate the three-body contribution by subtracting out the pairwise terms
	\begin{equation}
		\label{eq:F123}
		\mathcal{F}_{123} = \mathcal{F}^{(3)} - \sum_{i<j} \mathcal{F}^{(2)}_{ij}\,.
	\end{equation}
	Equation~(\ref{eq:F123}) captures collective interactions that arise only in the presence of all three colloids, and similar to the two-body interaction, it is sensitive to geometric arrangement and boundary conditions.
	
	\subsection*{Scattering and Translation Matrices}

	Since the interface fluctuations are described by the capillary-wave Hamiltonian, the corresponding modified Helmholtz equation defines the basis of incident and scattered modes. These modes can be used to create the single-particle scattering properties and determine the $\mathbb{T}$-matrix of a colloid. 
	Therefore, the Dirichlet T-matrix elements for a single spherical colloid of radius $R_i$ are determined by the boundary conditions imposed on the capillary wave field~\cite{nfr2013sct}. They are given in terms of modified Bessel functions as
	\begin{equation}
		\label{eq:tmatrix}
		\mathbb{T}^i_m = -\frac{I_m(R_i/\lambda_\mathrm{c})}{K_m(R_i/\lambda_\mathrm{c})} \,.
	\end{equation}
	In this expression, $m$ labels the multipole mode, while $\lambda_\mathrm{c}=\sqrt{\sigma/(\Delta\rho\,g)}$ defines the characteristic capillary length set by the surface tension $\sigma$, the density difference $\Delta\rho$ between the fluids, and the gravitational acceleration $g$. The corresponding translation matrix relating colloids $i$ and $j$ is given by
	\begin{equation}
		\label{eq:umatrix}
		\mathbb{U}^{ij}_{mm'} = (-1)^{m'} e^{i(m - m')\theta_{ij}} K_{m - m'}(d_{ij}/\lambda_\mathrm{c}) \,,
	\end{equation}
	where $d_{ij}$ is the center-to-center distance between the colloids and $\theta_{ij}$ is the angle between their coordinate systems.
	
	\subsection*{Boundary Conditions and Degrees of Freedom}
	
	We consider three distinct fluctuation scenarios for the colloids at the interface, each corresponding to different boundary conditions:
	
	\begin{description}
		\item [\bf scenario a] -- {\em fixed colloids}: Particles are trapped at the fluid interface and cannot bob or tilt. All T-matrix elements are retained in this scenario, and $\widetilde{\mathbb{T}} = \mathbb{T}$.
		
		\item[\bf scenario b] -- {\em bobbing colloids}: Colloids are allowed to fluctuate vertically but cannot tilt. In this case, the monopole component ($m = 0$) of the T-matrix $\mathbb{T}$ is set to zero.
		
		\item[\bf scenario c] -- {\it bobbing and tilting colloids}: Colloids are allowed to move both vertically and rotationally. The monopole and dipole components ($m = 0, \pm 1$) are set to zero in the T-matrix $\mathbb{T}$.
	\end{description}
	
	This procedure ensures that only physically allowed capillary modes contribute to the interaction energy in each scenario \cite{nfr2013sct}.
	
	\subsection*{Asymptotic Energy Expressions}
	
	At large separations ($d \gg R$), one can find analytical expressions for the interaction energy between spherical colloids by using the asymptotic forms of Bessel functions. The asymptotic expressions reveals the scaling behavior of the interaction and also help us to validate numerical results in Sec.~\ref{sec:numerical} against the large distance relations.
	
	\subsubsection*{Two-body interaction:}
	In order to estimate the Casimir interaction between two colloidal particles at the interface, we find large distance \(d_{12} \gg R_1, R_2\) expressions for the scenarios (a)--(c). At such large separations, the interaction energy comes from the lowest multipoles in the multipole expansion, which results in a simple scaling behavior of the interaction energy~\cite{nfr2013sct}.
	
	In scenario (a), the colloids are fixed, and the leading term of the energy expansion is associated with the monopole mode. The monopole term has a logarithmic behavior, due to the asymptotic behavior of the Bessel functions $I_0(R_i/\lambda_\mathrm{c})$ and $K_0(R_i/\lambda_\mathrm{c})$. This leads to a logarithmic relationship with the distance between the colloidal particles; we call it the coupling parameter \(g_{ij}\),
	\begin{equation}
	\label{eq:g}
		g_{ij} = \frac{\ln(2\lambda_\mathrm{c}/d_{ij})}{\sqrt{\ln(2\lambda_\mathrm{c}/R_i)\ln(2\lambda_\mathrm{c}/R_j)}} \,.
	\end{equation}
	In scenario (b), the colloids are allowed to bob. The monopole contribution in this scenario disappears due to the resulting boundary condition. Therefore, the dipole term becomes the dominant term, and consequently, the interaction decreases directly with the inverse of distance. Similarly, in scenario (c), both bobbing and tilting motions are permitted. Therefore, the first two terms in the expansion, i.e., monopole and dipole terms disappear, and the quadrupole mode is the leading contribution. This results in a faster energy decay with the inverse of distance.
	
	We summarized the asymptotic results for two spherical colloids obtained in Ref.~\cite{nfr2013sct} in Table~\ref{tab:2p-asym}. One can observe that  
	the long-range behavior of the energy changes from logarithmic in the fixed case to quickly decreasing with the inverse of distance when we add more degrees of freedom. 
	This comparison illustrates how colloids' movement alters the behavior of 
	fluctuation-induced forces at fluid interfaces.
	
	\begin{table}[h!]
		\label{tab:2p-asym}
		\centering
		\begin{tabular}{|c|c|}
			\hline
			Scenario & $\mathcal{F}^{(2)}/k_\mathrm{B}T$ at $d \gg R$ \\
			\hline 
			(a)  &$\frac{1}{2}\ln(1 - g_{12}^2)$ \\ \hline
			(b)  & $-\frac{R_1^2 R_2^2}{d_{12}^4}$ \\ \hline
			(c)  & $-9 \frac{R_1^4 R_2^4}{d_{12}^8}$ \\ \hline
		\end{tabular}
		\caption{Asymptotic forms of the two-body Casimir interaction for various colloids fluctuations scenarios, adapted from Ref.~\cite{nfr2013sct}.}
		\label{tab:2p-asym}
	\end{table}

	\subsubsection*{Three-body interaction:}
	For the three-body Casimir interaction $\mathcal{F}_{123}$, we can derive similar forms for large separations between the colloidal particles. Similar to the two-body interaction, the asymptotic energy is mainly determined by the lowest multipole moments for each scenario. For fixed colloids at the interface (scenario a), as explained above, the leading term of the multipole expansion has a logarithmic behavior. This would lead to non-additive coupling between the monopole terms of the three particles, see Table~\ref{tab:3p-asym}. 
	
	In the bobbing scenario (scenario b), since the main contribution comes from the dipole moment, we observe a decay that is much shorter-ranged than the fixed case. The strength of the interaction depends on the particle sizes. When both bobbing and tilting are allowed (scenario c), removing both monopole and dipole moments makes the quadrupole term dominant. This leads to a faster decay with the inverse of distance and similar to bobbing case a dependence on the size of the colloids.
	
	The asymptotic expressions for all three scenarios, which are obtained in Ref.~\cite{nfr2013sct} are shown in Table~\ref{tab:3p-asym}. These findings show how the degrees of freedom of the colloids influence both the range and the scaling behavior of the three-body Casimir interaction. Similar to the two-body interaction, they also act as analytical references for evaluating numerical results at intermediate distances.
	\begin{table}[h!]
		\centering
		\begin{tabular}{|c|c|}
			\hline
			Scenario & $\mathcal{F}^{(3)}/k_\mathrm{B}T$ at $d_{ij} \gg R$ \\
			\hline
			(a)  & $\frac{1}{2}\ln\left[\frac{1 - (g_{12}^2 + g_{13}^2 + g_{23}^2 - 2g_{12}g_{13}g_{23})}{(1 - g_{12}^2)(1 - g_{13}^2)(1 - g_{23}^2)}\right]$ \\
			\hline
			(b) & $-\sum\limits_{\text{cyc}}\frac{R_i^2 R_j^4 R_k^2}{d_{ij}^4 d_{jk}^4}$ \\
			\hline
			(c) & $-81\sum\limits_{\text{cyc}}\frac{R_i^4 R_j^8 R_k^4}{d_{ij}^8 d_{jk}^8}$ \\
			\hline
		\end{tabular}
		\caption{Asymptotic forms of the three-body Casimir interaction in different scenarios, adapted from Ref.~\cite{nfr2013sct}.}
		\label{tab:3p-asym}
	\end{table}

	\section{Numerical Calculation}
	\label{sec:numerical}
	
	In this section, we outline the numerical algorithm that we used for calculating the 
	Casimir energy between two or three spherical Janus colloids that are 
	trapped at the air-water interface. We consider nanometer-sized colloids at the air-water interface. For such interfaces the characteristic capillary length is about  $\lambda_\mathrm{c}/R\sim 10^{6}$, where $R$ is the colloid radius.
	
	The Casimir interaction energy between two or three colloids 
	is calculated by using Eq.~(\ref{eq:F2}) for the pairwise and 
	Eq.~(\ref{eq:F3}) for the three-body interactions. 
	For a separation $d_{ij}$ between two colloids $i$ and $j$, 
	we calculate the $f_{ij}$-block in the energy expressions, 
	Eq.~(\ref{eq:f-elements}), numerically.
	The maximum number of multipoles kept in the calculations, denoted by $N$, 
	determines the size of the scattering ($\mathbb{T}$) matrix, translation ($\mathbb U$) matrix, and consequently the size of the $f$-blocks.
	
	We determine the maximum number of multipoles $N$  required to find the correct value 
	of the energy as follows:
	
	We initiate the calculation with the lowest-order multipole allowed by each scenario. 
	In scenario (a), the lowest mode is $m = 0$ (monopole); in scenario (b), $m = \pm1$ (dipole); and in scenario (c), $m = \pm2$ (quadrupole). 
	The $\mathbb{T}$- and $\mathbb{U}$-matrices corresponding to the lowest modes take different forms depending on the scenario. 
	In scenario (a), they are scalars. In scenario (b), which involves bobbing colloids, the matrices are $3 \times 3$, with the $m = 0$ components of the T-matrix vanishing. 
	In scenario (c), where colloids exhibit both bobbing and tilting, the matrices are $5 \times 5$, and the T-matrix elements corresponding to $m = 0, \pm1$ vanish.
	The corresponding energy at this level is denoted by $\mathcal{F}_1^{(i)}$, 
	where $i=2,3$ refers to pairwise or three-body interactions.

	We then extend the calculation by including the next set of multipole moments appropriate to each scenario and evaluate the corresponding interaction energy. 
	Specifically, in scenario (a) we add $m = \pm1$; in scenario (b), $m = \pm2$; and in scenario (c), $m = \pm3$. 
	At this level of truncation, the T- and U-matrices for each colloid become $3 \times 3$, $5 \times 5$, and $7 \times 7$ for scenarios (a), (b), and (c), respectively. 
	The updated interaction energy is denoted by $\mathcal{F}_2^{(i)}$, where $i = 2, 3$ corresponds to the pairwise and three-body cases, as before.

	We then define the energy difference between successive truncation as
	\[
	\Delta E = |\mathcal{F}_2^{(i)} - \mathcal{F}_1^{(i)}|\,.
	\]
	We introduce a convergence tolerance $\varepsilon$, typically ranging between $10^{-8}$ and $10^{-4}$. If $\Delta E < \varepsilon$, the 
	calculation is deemed converged. Otherwise, we update $\mathcal{F}_1^{(i)} \leftarrow \mathcal{F}_2^{(i)}$ and proceed by including the next-
	order multipoles for each scenario. This procedure is iterated, incrementally incorporating higher-order terms until convergence is achieved within the prescribed tolerance.
	\begin{figure}
		\centering
		\includegraphics[width=0.7\columnwidth]{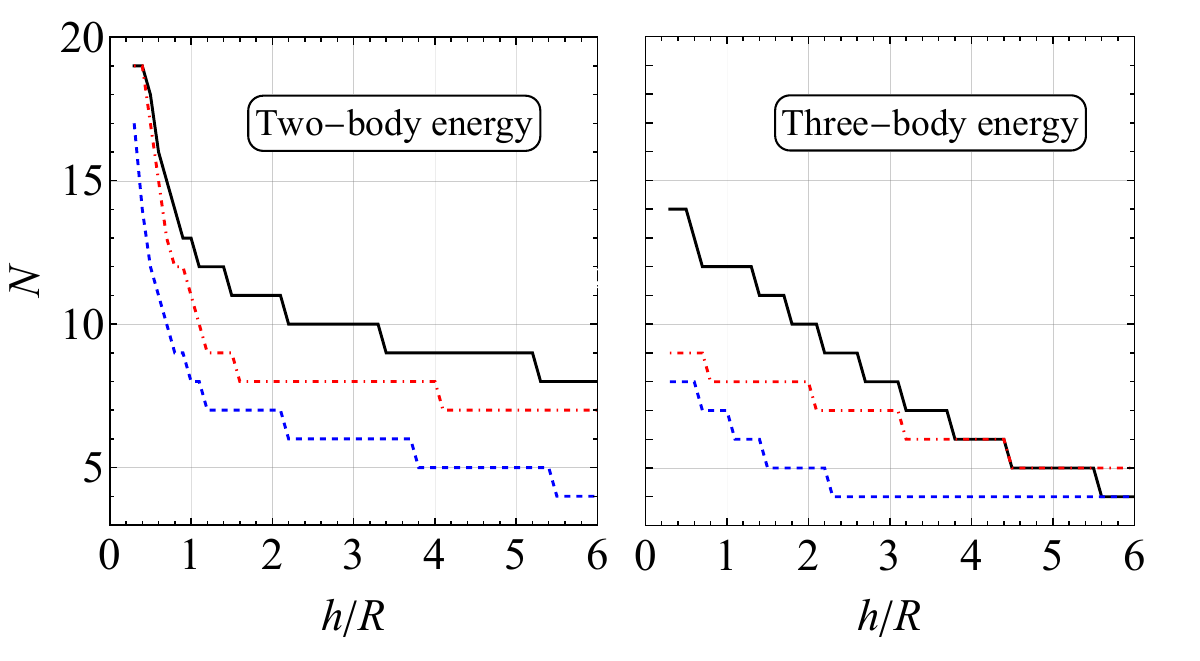}
		\caption{
			Number of multipoles $N$ required to achieve convergence of the Casimir energy between identical colloids of radius $R$, plotted as a function of the dimensionless surface-to-surface distance $h/R$. 
			The left panel shows $N$ for the two-body interaction $\mathcal{F}^{(2)}$, while the right panel presents $N$ for the three-body contribution $\mathcal{F}_{123}$. 
			In both plots, the dashed blue, dot-dashed red, and solid black curves correspond to scenarios (a), (b), and (c), respectively.
		}
		\label{fig:1}
	\end{figure}

	Figure~\ref{fig:1} presents the number of multipoles $N$ required for convergence of the Casimir energy between frozen, monodisperse spherical colloids as a function of the scaled surface-to-surface gap $h/R$, defined by
	\[
	h = d - 2R\,.
	\]

	As illustrated in both panels, the number of multipoles required to achieve convergence in the Casimir interaction energy increases significantly as the colloids are brought closer together. This behavior is consistent with the fact that shorter interparticle distances demand a more accurate resolution of the near-field interactions, which are increasingly sensitive to higher-order multipole contributions. In other words, as the colloids approach each other, the complexity of  field-mediated interactions increases, requiring the inclusion of a greater number of partial waves to capture the interaction.
	
	Moreover, we observe that the required number of multipoles strongly depends on the mechanical constraints imposed on the colloids at the interface. Specifically, in scenario (a), where the colloids are fixed in position and orientation, convergence is noticeably faster, and a smaller number of multipoles is needed compared to scenarios (b) and (c), where the colloids are allowed to bob or bob-and-tilt, respectively. This can be attributed to the fact that, in scenario (a), the interaction is dominated by the monopole contribution, which exhibits a strong logarithmic behavior and thus governs the overall scaling of the Casimir energy. In contrast, when the colloids are free to fluctuate (scenarios b and c), the monopole mode is entirely eliminated due to the boundary conditions associated with these degrees of freedom. As a result, the leading contribution arises from higher-order modes (dipole, quadrupole, etc.), which decay more rapidly with distance and lead to slower convergence.
	
	Another important observation concerns the sensitivity of the results to the numerical tolerance parameter $\varepsilon$ used to define convergence. As expected, reducing the tolerance leads to a stricter convergence criterion, thereby requiring more multipole orders to be included in the calculation. This effect is particularly pronounced at short separations, where small deviations in the energy are more difficult to control due to the rapid growth of higher-order terms.
	
	Interestingly, we also find that, for a fixed value of $\varepsilon$, the number of multipoles needed to reach convergence is generally smaller for the three-body contribution $\mathcal{F}_{123}$ than for the pairwise energy $\mathcal{F}^{(2)}$. This behavior might seem counterintuitive at first, as the three-body interaction is structurally more complex. However, the coupling in the three-body configurations may suppress certain contributions, which can accelerate convergence in practice.
	
	These results collectively highlight the subtle interplay between geometry, boundary conditions, and numerical precision in determining the convergence behavior of Casimir energies in polydisperse colloidal systems.

	Furthermore, we verified that this convergence behavior also holds for polydisperse configurations. While the general trend of requiring more multipoles for smaller gaps remains unchanged, we note that for a fixed separation, a large radius ratio ($r \gtrsim 4$) may require a modest increase in the number of multipoles (typically one or two) to reach the same convergence tolerance. This is physically expected, as the enhanced geometric asymmetry of the configuration can strengthen the contribution of higher-order modes.
	
	\section{Results and Discussions}
	\label{sec:discussion}
	
	\subsection{Two-body interaction}
	
	To investigate the effect of polydispersity on the Casimir interaction between two colloidal particles, we consider a heterogeneous configuration in which the colloids have unequal radii. We compare the Casimir free energy of this system with a 
	homogeneous configuration in which both colloids have the same radius. To quantify the impact of polydispersity, we introduce a dimensionless energy deviation parameter, $\Delta_2$, which represents the deviation of the interaction energy in the heterogeneous configuration from that of the homogeneous reference system
	\begin{equation}
		\label{eq:2part-disparity}
		\Delta_2 = \frac{\mathcal{F}^{(2)}(R_<, R_>, h_{12})}{\mathcal{F}^{(2)}(R_<, R_<, h_{12})} - 1\,,
	\end{equation}
	where $\mathcal{F}^{(2)}$ given by Eq.~(\ref{eq:F2}) denotes the Casimir interaction energy between two colloids, and $R_< < R_>$ are the radii of the smaller and larger colloid, respectively. For convenience, we define the radius ratio
	$$
	r = \frac{R_>}{R_<}\,.
	$$
	Throughout this analysis, the surface-to-surface distance $h_{12}$ between the colloids is held fixed. By varying $r$ while keeping $h_{12}$ constant at $h_{12}/r=5$, we isolate the effect of polydispersity on the interaction energy and study how deviations from size homogeneity modify the Casimir force.
	\begin{figure}[ht]
		\centering
		\includegraphics[width=0.6\columnwidth]{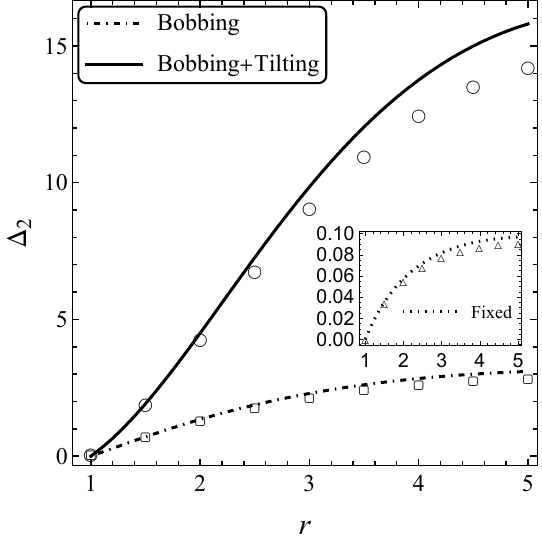}
\caption{Two-body Casimir energy deviation parameter $\Delta_{2}$ as a function of the radius ratio $r=R_{>}/R_{<}$ for three different boundary conditions, at a fixed surface-to-surface distance of $h/R_{<}=5$. The full numerical results ({symbols}) are compared against the large-distance asymptotic theory ({lines}) from Table~\ref{tab:2p-asym}. The main plot displays the results for bobbing (squares) and bobbing+tilting (circles). The inset shows the much smaller deviation for the fixed case (triangles). }
		\label{fig:2}
	\end{figure}

	Figure~\ref{fig:2} depicts the behavior of $\Delta_2$ as a function of $r$ for three scenarios (a), (b), and (c). We see that in all three cases, the deviation parameter $\Delta_2$ increases with $r$, which means that as the size difference between the colloids grows, the Casimir interaction energy deviates more from  the  monodisperse case, which is also showing excellent agreement between the full numerical calculations (symbols) and the large-distance asymptotic theory (lines).

	As it is illustrated in Fig.~\ref{fig:2}, the bobbing+tilting colloids (scenario c) exhibit 
	higher sensitivity to the polydispersity of colloids. As shown by the circles, the energy deviation parameter $\Delta_2$ exceeds 10 when the radius ratio reaches $r \sim 5$. The bobbing-only colloids (squares) shows a moderate increase in $\Delta_2$ with $r$, remaining below approximately 3 over the same range.
	The fixed colloid configuration (the triangles in the inset) shows the smallest values of $\Delta_2$, with the deviation remaining below 1\% even for $r\sim5$. These curves indicate that polydispersity has a stronger impact on the Casimir interaction when the colloids are allowed to fluctuate with more degrees of freedom.  This enhanced sensitivity means that  higher-order terms such as dipole and quadrapole can enhance the effect of polydispersity, depending on the specific colloids fluctuation scenario.
	At first glance, this may seem counterintuitive, since fixed boundary conditions typically preserve strong monopole contribution with the dominant logarithmic behavior. However, we can understand this better in the context of polydispersity and long-distance regime. In the fixed colloids case (scenario a) the dependence on $r$ appears in the denominator of the coupling 
	parameter $g$, see Eq.~(\ref{eq:g}) which would not have a determining effect on the two-body Casimir energy but in the 
	scenarios (b) and (c) the Casimir energy scales with power-laws of the colloids radii 
	whcih has much stronger effect, see Tab.~\ref{tab:2p-asym}. Therefore, while the overall Casimir energy may be large in the fixed case, the relative sensitivity to radius mismatch is reduced.
	Consequently, the system exhibits a kind of amplified sensitivity to polydispersity when additional degrees of freedom are allowed in colloids fluctuations.
	
	The physical origin of this amplified sensitivity can be understood by considering how the colloid's mechanical freedom alters its coupling to the spectrum of capillary wave fluctuations. For a fixed colloid, the particle is rigidly constrained and thus interacts with all fluctuation modes, including the powerful, long-wavelength monopole mode. This strong but geometry-insensitive interaction dominates, masking more subtle effects.
	
	However, when a colloid is allowed to bob, its vertical degree of freedom allows it to move in sync with the symmetric, long-wavelength monopole fluctuations. By moving with the wave, the colloid effectively decouples from this mode, rendering it invisible to this part of the fluctuation spectrum. This decoupling acts as a physical filter. Similarly, allowing the colloid to also tilt lets it decouple from the dipole modes.
	
	By filtering out these dominant, lower-order modes, the interaction is forced to be governed exclusively by the remaining higher-order, short-wavelength modes (e.g., quadrupole). These modes are inherently more sensitive to the precise geometric details of the particle's surface. The ``amplified sensitivity'' is therefore a direct consequence of this physical decoupling mechanism, which unmasks the latent, extreme geometry-dependence of the higher-order interactions.
	
	\subsection{Three-body contribution}
	
	To investigate the influence of polydispersity on the three-body Casimir 
	interaction, we consider a linear, symmetric configuration composed of three 
	spherical colloids trapped at the air–water interface. In this symmetric setup, 
	the surface-to-surface distances between adjacent colloids are equal, $h_{12} = 
	h_{23}$, with colloid 2 located at the center, between colloids 1 and 3, thus $h_{13}=2(h_{12}+R_2)$. We 
	investigate the effect of polydispersity by assigning a larger radius to the 
	middle colloid, such that $R_2 = R_>$ and $R_1 = R_3 = R_<$, where $R_>$ and 
	$R_<$ denote the larger and smaller colloid radii, respectively.
	
	To quantify the effect of the polydispersity on the three-body Casimir interaction, we define a dimensionless deviation parameter $\Delta_{123}$, analogous to the two-body case, as
	\begin{equation}
		\label{eq:3part-disparity}
		\Delta_{123} = \frac{\mathcal{F}_{123}(R_<, R_>, R_<; d_{12}, d_{23}, 
			d_{13})}{\mathcal{F}_{123}(R_<, R_<, R_<; d_{12}, d_{23}, d_{13})} - 1\,,
	\end{equation}
	where $\mathcal{F}_{123}$ denotes the three-body Casimir contribution beyond pairwise additivity, as defined in Eq.~(\ref{eq:F123}). The distances between the colloids satisfy $d_{12} = d_{23} = d$ and $d_{13} = 2d$, ensuring the spatial symmetry of the configuration. Again, the radius ratio $r = R_>/R_<$ is the polydispersity parameter, with $r = 1$ corresponding to the monodisperse case.
	
	This framework allows us to study the effect of polydispersity on the three-body interaction by 
	changing the radius of larger central colloid, while keeping geometric symmetry unchanged. 
	As we will demonstrate, the behavior of $\Delta_{123}$ as a function of the radius ratio $r=R_>/R_<$ 	
	reveals a complex dependence on the colloids' degrees of freedom and separation distance, exhibiting both suppressive and enhancing effects.

\begin{figure}[ht]
		\centering
		\includegraphics[width=0.6\columnwidth]{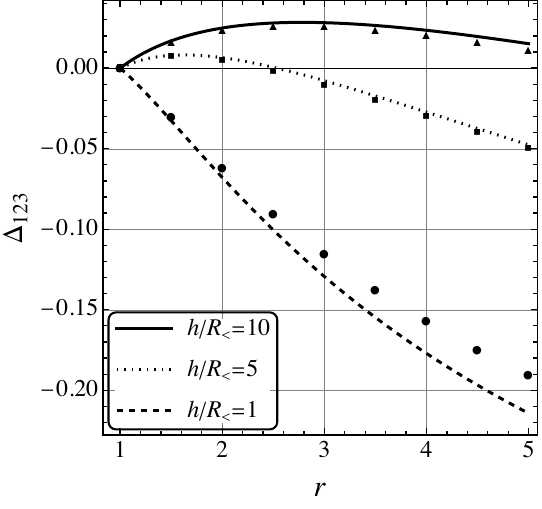}
\caption{Three-body deviation parameter $\Delta_{123}$ versus the radius ratio $r=R_{>}/R_{<}$ for fixed colloids (scenario a). The central colloid has radius $R_{>}$ and the outer ones have radius $R_{<}$. Lines show the large-distance asymptotic prediction from Table.~\ref{tab:3p-asym}, while symbols represent the full numerical calculation. Triangles, squares, and circles correspond to separations of $h/R_{<}=10,$ 5, and 1, respectively.}
		\label{fig:3}
	\end{figure}

\begin{figure}[ht]
		\centering
		\includegraphics[width=0.6\columnwidth]{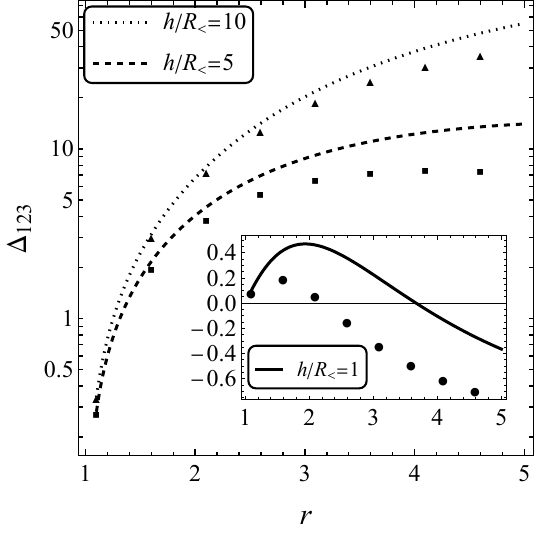}
\caption{Three-body deviation parameter $\Delta_{123}$ versus $r=R_{>}/R_{<}$ for {bobbing colloids} (scenario b). Lines are the asymptotic forms from Table~II, while symbols denote the full numerical results (triangles for $h/R_{<}=10$, squares for $h/R_{<}=5$, and circles for $h/R_{<}=1$). The main plot is presented on a logarithmic scale, while the inset uses a linear scale to detail the behavior at the shortest separation, $h/R_{<}=1$.}
		\label{fig:4}
	\end{figure}

\begin{figure}[ht]
		\centering
		\includegraphics[width=0.6\columnwidth]{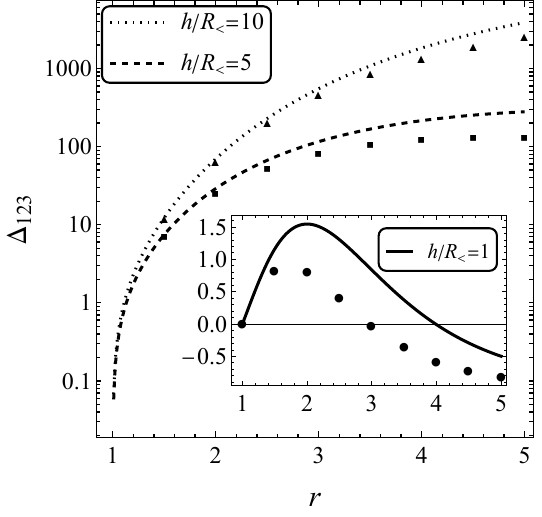}
\caption{Three-body deviation parameter $\Delta_{123}$ versus $r=R_{>}/R_{<}$ for bobbing and tilting colloids (scenario c). Lines are the asymptotic forms from Table~II, while symbols denote the full numerical results (triangles for $h/R_{<}=10$, squares for $h/R_{<}=5$, and circles for $h/R_{<}=1$). The main plot is presented on a logarithmic scale, while the inset uses a linear scale to focus on the behavior at the shortest separation, $h/R_{<}=1$.}		\label{fig:5}
	\end{figure}

Before discussing the results for each scenario in detail, it is important to note that our full numerical calculations show excellent agreement with the asymptotic 
predictions in their respective regimes of validity. As can be seen in Figures~\ref{fig:3}-\ref{fig:5}, while there are expected deviations at short separations, the 
numerical data (symbols) consistently follow the trend predicted by the asymptotic theory (lines) at large distances. This strong correspondence validates the 
robustness of our numerical results.
	
For fixed colloids (scenario a), the effect of polydispersity on the three-body interaction is highly dependent on separation, 
as shown in Fig.~\ref{fig:3}. At the shortest separation ($h/R_{<}=1$), the effect is purely {suppressive}, 
with $\Delta_{123}$ becoming increasingly negative as the size ratio $r$ grows. At an intermediate separation ($h/R_{<}=5$), 
the interaction becomes {competitive}, exhibiting a weak enhancement at small $r$ before crossing over to a suppressive regime. 
Interestingly, at the largest separation ($h/R_{<}=10$), the effect reverses completely and becomes weakly but consistently enhancing. 
This behavior reveals that even for fixed colloids, the interplay between geometry and separation can qualitatively alter the nature of the many-body force.

Allowing the colloids to bob (scenario b) leads to a predominantly {enhancing} effect, as presented in Fig~\ref{fig:4}. 
At large and intermediate separations ($h/R_{<}=10$ and $h/R_{<}=5$), the deviation parameter $\Delta_{123}$ is purely positive and grows monotonically with the radius ratio $r$. 
However, at the shortest separation ($h/R_{<}=1$), the interaction becomes {competitive}. As shown in the inset of Figure~4, the effect is weakly enhancing for small $r$ before crossing over to become suppressive at $r \approx 2.2$. 
This demonstrates that while the vertical degree of freedom generally promotes enhancement, a complex competitive behavior emerges in the near-field.	
	
The highest sensitivity to polydispersity is observed in scenario (c), where colloids are allowed to both bob and tilt. The results, shown in Fig.~\ref{fig:5}, are even 
stronger than the bobbing-only case. At large and intermediate separations ($h/R_{<}=10$ and $h/R_{<}=5$), the effect is purely and strongly {enhancing}, with the 
deviation $\Delta_{123}$ growing by multiple orders of magnitude. This confirms a remarkable ``amplified sensitivity.'' In contrast, at the shortest separation ($h/
R_{<}=1$), a {competitive} behavior emerges: the interaction is weakly enhancing at small $r$ before becoming suppressive at $r \approx 3$, as shown in the inset. 
This extreme long-range enhancement arises because the suppression of both monopole and dipole modes leaves the interaction to be governed exclusively by 
higher-order, geometry-sensitive multipoles.

The rich competitive behavior observed in our results can be understood as a competition between two distinct physical effects governing the near-field and far-field regimes.

At {short separations}, especially for a large radius ratio ($r \gg 1$), the geometric configuration approaches that of two small colloids near a large, quasi-planar wall. 
The strong and uniform boundary condition imposed by this ``wall'' alters the mediating capillary wave modes in a way that weakens the non-additive three-body 
contribution compared to the reference system of three small colloids. This leads to the {suppressive effect} ($\Delta_{123} < 0$) observed in the near-field for most 
scenarios.

Conversely, at {large separations}, the interaction becomes ``purified.'' Higher-order multipole contributions, which are significant in the near-field, decay away. For 
mobile colloids, the dominant lower-order modes are also physically filtered out by the particles' motion. The interaction is thus governed exclusively by a single, 
highly geometry-sensitive mode (e.g., quadrupole). In this ``pure'' far-field regime, the mode's strong intrinsic dependence on particle radii is fully unmasked. This is 
why the \textit{relative} effect of polydispersity becomes more pronounced as the colloids move farther apart, leading to the dramatic {enhancement} seen in our 
results. The observed crossover phenomena are therefore a manifestation of the competition between these near-field geometric suppression and far-field modal 
purification effects.	

The numerical data used to generate Figs.~\ref{fig:1}–\ref{fig:5} can be accessed at the Zenodo repository~\cite{zenododata}.
	
	\section{Summary and Conclusions}
	\label{sec:summary}

In this work, we have investigated the role of polydispersity in fluctuation-induced (Casimir-like) interactions between colloids trapped at a fluid interface. By moving 
beyond the common assumption of monodispersity, our findings reveal that size asymmetry is not merely a minor perturbation but acts as a {qualitative tuner} of the 
interaction, capable of inducing both suppression and enhancement of the many-body forces.

Our central finding is a clear dichotomy in the system's response, which is governed by the mechanical degrees of freedom of the colloids. For fixed colloids, the 
effect is predominantly suppressive, though a complex dependence on separation is observed. In contrast, when vertical and rotational fluctuations are 
permitted, the nature of the force is fundamentally altered, becoming either purely enhancing or competitive, depending on the separation. This demonstrates that 
colloidal mobility acts as a {switch}, qualitatively redefining the impact of polydispersity on the interaction landscape.

The physical mechanism underlying this switch is the selective suppression of multipole interaction modes. For fixed colloids, the interaction involves a complex 
interplay of all modes. However, allowing the colloids to move filters out the dominant, lower-order modes. This ``unmasking'' of higher-order, geometry-sensitive 
multipoles results in the amplified sensitivity observed at long ranges.

Indeed, our findings offer significant new insights into the self-assembly of polydisperse colloids at interfaces. The discovery that the effect of polydispersity can be qualitatively switched from suppressive to enhancing has profound implications. For instance:
\begin{itemize}
    \item[-] {Tunable Aggregation:} The crossover behaviors can create complex energy landscapes with kinetic barriers and potential wells, leading to size-selective aggregation where clusters of a specific size ratio are favored.
    \item[-] {Directed Pattern Formation:} Since the forces are exquisitely sensitive to the size of the central particle, polydispersity could be used as a tool to direct the assembly of non-trivial patterns, such as specific linear chains or asymmetric clusters.
    \item[-] {Stability of Colloidal Crystals:} A small degree of polydispersity could significantly alter the stability of different lattice structures by introducing a complex network of suppressive and enhancing many-body forces.
\end{itemize}

Finally, our findings also highlight the profound challenges and opportunities for experimental verification. The predicted long-range enhancement, while representing a dramatic relative amplification, corresponds to an absolute interaction energy, $\mathcal{F}_{123}$, that is expected to fall well below the thermal energy scale, $k_BT$. A direct force measurement of this long-range effect is therefore likely beyond the reach of current experimental capabilities. Nevertheless, the key qualitative predictions, such as the crossover phenomena at shorter, more accessible separations, could potentially be tested, perhaps through statistical analysis of particle trajectories. Our theoretical framework thus lays essential groundwork for interpreting future experiments in these rich and complex systems.

	\bibliography{mybib}{}

\begin{thebibliography}{34}%
\makeatletter
\providecommand \@ifxundefined [1]{%
 \@ifx{#1\undefined}
}%
\providecommand \@ifnum [1]{%
 \ifnum #1\expandafter \@firstoftwo
 \else \expandafter \@secondoftwo
 \fi
}%
\providecommand \@ifx [1]{%
 \ifx #1\expandafter \@firstoftwo
 \else \expandafter \@secondoftwo
 \fi
}%
\providecommand \natexlab [1]{#1}%
\providecommand \enquote  [1]{``#1''}%
\providecommand \bibnamefont  [1]{#1}%
\providecommand \bibfnamefont [1]{#1}%
\providecommand \citenamefont [1]{#1}%
\providecommand \href@noop [0]{\@secondoftwo}%
\providecommand \href [0]{\begingroup \@sanitize@url \@href}%
\providecommand \@href[1]{\@@startlink{#1}\@@href}%
\providecommand \@@href[1]{\endgroup#1\@@endlink}%
\providecommand \@sanitize@url [0]{\catcode `\\12\catcode `\$12\catcode
  `\&12\catcode `\#12\catcode `\^12\catcode `\_12\catcode `\%12\relax}%
\providecommand \@@startlink[1]{}%
\providecommand \@@endlink[0]{}%
\providecommand \url  [0]{\begingroup\@sanitize@url \@url }%
\providecommand \@url [1]{\endgroup\@href {#1}{\urlprefix }}%
\providecommand \urlprefix  [0]{URL }%
\providecommand \Eprint [0]{\href }%
\providecommand \doibase [0]{https://doi.org/}%
\providecommand \selectlanguage [0]{\@gobble}%
\providecommand \bibinfo  [0]{\@secondoftwo}%
\providecommand \bibfield  [0]{\@secondoftwo}%
\providecommand \translation [1]{[#1]}%
\providecommand \BibitemOpen [0]{}%
\providecommand \bibitemStop [0]{}%
\providecommand \bibitemNoStop [0]{.\EOS\space}%
\providecommand \EOS [0]{\spacefactor3000\relax}%
\providecommand \BibitemShut  [1]{\csname bibitem#1\endcsname}%
\let\auto@bib@innerbib\@empty
\bibitem [{\citenamefont {Binks}\ and\ \citenamefont
  {Horozov}(2006)}]{binks2006colloidal}%
  \BibitemOpen
  \bibfield  {author} {\bibinfo {author} {\bibfnamefont {B.~P.}\ \bibnamefont
  {Binks}}\ and\ \bibinfo {author} {\bibfnamefont {T.~S.}\ \bibnamefont
  {Horozov}},\ }\href@noop {} {\emph {\bibinfo {title} {Colloidal Particles at
  Liquid Interfaces}}}\ (\bibinfo  {publisher} {Cambridge University Press},\
  \bibinfo {year} {2006})\BibitemShut {NoStop}%
\bibitem [{\citenamefont {Stamou}\ \emph {et~al.}(2000)\citenamefont {Stamou},
  \citenamefont {Duschl},\ and\ \citenamefont {Johannsmann}}]{stamou2000long}%
  \BibitemOpen
  \bibfield  {author} {\bibinfo {author} {\bibfnamefont {D.}~\bibnamefont
  {Stamou}}, \bibinfo {author} {\bibfnamefont {C.}~\bibnamefont {Duschl}},\
  and\ \bibinfo {author} {\bibfnamefont {D.}~\bibnamefont {Johannsmann}},\
  }\href {https://doi.org/10.1103/PhysRevE.62.5263} {\bibfield  {journal}
  {\bibinfo  {journal} {Phys. Rev. E}\ }\textbf {\bibinfo {volume} {62}},\
  \bibinfo {pages} {5263} (\bibinfo {year} {2000})}\BibitemShut {NoStop}%
\bibitem [{\citenamefont {Vella}\ and\ \citenamefont
  {Mahadevan}(2005)}]{vella2005cheerios}%
  \BibitemOpen
  \bibfield  {author} {\bibinfo {author} {\bibfnamefont {D.}~\bibnamefont
  {Vella}}\ and\ \bibinfo {author} {\bibfnamefont {L.}~\bibnamefont
  {Mahadevan}},\ }\href {https://doi.org/10.1119/1.1898523} {\bibfield
  {journal} {\bibinfo  {journal} {American Journal of Physics}\ }\textbf
  {\bibinfo {volume} {73}},\ \bibinfo {pages} {817} (\bibinfo {year}
  {2005})}\BibitemShut {NoStop}%
\bibitem [{\citenamefont {Oettel}\ \emph {et~al.}(2005)\citenamefont {Oettel},
  \citenamefont {Dom\'{\i}nguez},\ and\ \citenamefont
  {Dietrich}}]{oettel2009effective}%
  \BibitemOpen
  \bibfield  {author} {\bibinfo {author} {\bibfnamefont {M.}~\bibnamefont
  {Oettel}}, \bibinfo {author} {\bibfnamefont {A.}~\bibnamefont
  {Dom\'{\i}nguez}},\ and\ \bibinfo {author} {\bibfnamefont {S.}~\bibnamefont
  {Dietrich}},\ }\href {https://doi.org/10.1103/PhysRevE.71.051401} {\bibfield
  {journal} {\bibinfo  {journal} {Phys. Rev. E}\ }\textbf {\bibinfo {volume}
  {71}},\ \bibinfo {pages} {051401} (\bibinfo {year} {2005})}\BibitemShut
  {NoStop}%
\bibitem [{\citenamefont {Lehle}\ and\ \citenamefont
  {Oettel}(2007)}]{oettel2007importance}%
  \BibitemOpen
  \bibfield  {author} {\bibinfo {author} {\bibfnamefont {H.}~\bibnamefont
  {Lehle}}\ and\ \bibinfo {author} {\bibfnamefont {M.}~\bibnamefont {Oettel}},\
  }\href {https://doi.org/10.1103/PhysRevE.75.011602} {\bibfield  {journal}
  {\bibinfo  {journal} {Phys. Rev. E}\ }\textbf {\bibinfo {volume} {75}},\
  \bibinfo {pages} {011602} (\bibinfo {year} {2007})}\BibitemShut {NoStop}%
\bibitem [{\citenamefont {Lehle}\ \emph {et~al.}(2006)\citenamefont {Lehle},
  \citenamefont {Oettel},\ and\ \citenamefont {Dietrich}}]{lehle2006effective}%
  \BibitemOpen
  \bibfield  {author} {\bibinfo {author} {\bibfnamefont {H.}~\bibnamefont
  {Lehle}}, \bibinfo {author} {\bibfnamefont {M.}~\bibnamefont {Oettel}},\ and\
  \bibinfo {author} {\bibfnamefont {S.}~\bibnamefont {Dietrich}},\ }\href
  {https://doi.org/10.1209/epl/i2006-10065-1} {\bibfield  {journal} {\bibinfo
  {journal} {Europhysics Letters}\ }\textbf {\bibinfo {volume} {75}},\ \bibinfo
  {pages} {174} (\bibinfo {year} {2006})}\BibitemShut {NoStop}%
\bibitem [{\citenamefont {Domínguez}\ \emph {et~al.}(2005)\citenamefont
  {Domínguez}, \citenamefont {Oettel},\ and\ \citenamefont
  {Dietrich}}]{dominguez2005capillary}%
  \BibitemOpen
  \bibfield  {author} {\bibinfo {author} {\bibfnamefont {A.}~\bibnamefont
  {Domínguez}}, \bibinfo {author} {\bibfnamefont {M.}~\bibnamefont {Oettel}},\
  and\ \bibinfo {author} {\bibfnamefont {S.}~\bibnamefont {Dietrich}},\ }\href
  {https://doi.org/10.1088/0953-8984/17/45/026} {\bibfield  {journal} {\bibinfo
   {journal} {Journal of Physics: Condensed Matter}\ }\textbf {\bibinfo
  {volume} {17}},\ \bibinfo {pages} {S3387} (\bibinfo {year}
  {2005})}\BibitemShut {NoStop}%
\bibitem [{\citenamefont {Domínguez}\ \emph {et~al.}(2007)\citenamefont
  {Domínguez}, \citenamefont {Oettel},\ and\ \citenamefont
  {Dietrich}}]{dominguez_2007}%
  \BibitemOpen
  \bibfield  {author} {\bibinfo {author} {\bibfnamefont {A.}~\bibnamefont
  {Domínguez}}, \bibinfo {author} {\bibfnamefont {M.}~\bibnamefont {Oettel}},\
  and\ \bibinfo {author} {\bibfnamefont {S.}~\bibnamefont {Dietrich}},\ }\href
  {https://doi.org/10.1209/0295-5075/77/68002} {\bibfield  {journal} {\bibinfo
  {journal} {Europhysics Letters}\ }\textbf {\bibinfo {volume} {77}},\ \bibinfo
  {pages} {68002} (\bibinfo {year} {2007})}\BibitemShut {NoStop}%
\bibitem [{\citenamefont {Dom\'{\i}nguez}\ \emph {et~al.}(2016)\citenamefont
  {Dom\'{\i}nguez}, \citenamefont {Malgaretti}, \citenamefont {Popescu},\ and\
  \citenamefont {Dietrich}}]{dominguez_2018}%
  \BibitemOpen
  \bibfield  {author} {\bibinfo {author} {\bibfnamefont {A.}~\bibnamefont
  {Dom\'{\i}nguez}}, \bibinfo {author} {\bibfnamefont {P.}~\bibnamefont
  {Malgaretti}}, \bibinfo {author} {\bibfnamefont {M.~N.}\ \bibnamefont
  {Popescu}},\ and\ \bibinfo {author} {\bibfnamefont {S.}~\bibnamefont
  {Dietrich}},\ }\href {https://doi.org/10.1103/PhysRevLett.116.078301}
  {\bibfield  {journal} {\bibinfo  {journal} {Phys. Rev. Lett.}\ }\textbf
  {\bibinfo {volume} {116}},\ \bibinfo {pages} {078301} (\bibinfo {year}
  {2016})}\BibitemShut {NoStop}%
\bibitem [{\citenamefont {Danov}\ and\ \citenamefont
  {Kralchevsky}(2010)}]{danov2010capillary}%
  \BibitemOpen
  \bibfield  {author} {\bibinfo {author} {\bibfnamefont {K.~D.}\ \bibnamefont
  {Danov}}\ and\ \bibinfo {author} {\bibfnamefont {P.~A.}\ \bibnamefont
  {Kralchevsky}},\ }\href
  {https://doi.org/https://doi.org/10.1016/j.cis.2010.01.010} {\bibfield
  {journal} {\bibinfo  {journal} {Advances in Colloid and Interface Science}\
  }\textbf {\bibinfo {volume} {154}},\ \bibinfo {pages} {91} (\bibinfo {year}
  {2010})}\BibitemShut {NoStop}%
\bibitem [{\citenamefont {Zeng}\ \emph {et~al.}(2012)\citenamefont {Zeng},
  \citenamefont {Brau}, \citenamefont {Davidovitch},\ and\ \citenamefont
  {Dinsmore}}]{zeng2012colloidal}%
  \BibitemOpen
  \bibfield  {author} {\bibinfo {author} {\bibfnamefont {C.}~\bibnamefont
  {Zeng}}, \bibinfo {author} {\bibfnamefont {F.}~\bibnamefont {Brau}}, \bibinfo
  {author} {\bibfnamefont {B.}~\bibnamefont {Davidovitch}},\ and\ \bibinfo
  {author} {\bibfnamefont {A.~D.}\ \bibnamefont {Dinsmore}},\ }\href
  {https://doi.org/10.1039/C2SM25871D} {\bibfield  {journal} {\bibinfo
  {journal} {Soft Matter}\ }\textbf {\bibinfo {volume} {8}},\ \bibinfo {pages}
  {8582} (\bibinfo {year} {2012})}\BibitemShut {NoStop}%
\bibitem [{\citenamefont {Colinet}\ and\ \citenamefont
  {Nepomnyashchy}(2010)}]{colinet_2010_pattern}%
  \BibitemOpen
  \bibfield  {author} {\bibinfo {author} {\bibfnamefont {P.}~\bibnamefont
  {Colinet}}\ and\ \bibinfo {author} {\bibfnamefont {A.}~\bibnamefont
  {Nepomnyashchy}},\ }\href@noop {} {\emph {\bibinfo {title} {Pattern Formation
  at Interfaces}}}\ (\bibinfo  {publisher} {Springer},\ \bibinfo {year}
  {2010})\BibitemShut {NoStop}%
\bibitem [{\citenamefont {Ghosh}\ \emph {et~al.}(2007)\citenamefont {Ghosh},
  \citenamefont {Fan},\ and\ \citenamefont {Stebe}}]{gosh2007}%
  \BibitemOpen
  \bibfield  {author} {\bibinfo {author} {\bibfnamefont {M.}~\bibnamefont
  {Ghosh}}, \bibinfo {author} {\bibfnamefont {F.}~\bibnamefont {Fan}},\ and\
  \bibinfo {author} {\bibfnamefont {K.~J.}\ \bibnamefont {Stebe}},\ }\href
  {https://doi.org/10.1021/la062150e} {\bibfield  {journal} {\bibinfo
  {journal} {Langmuir}\ }\textbf {\bibinfo {volume} {23}},\ \bibinfo {pages}
  {2180} (\bibinfo {year} {2007})}\BibitemShut {NoStop}%
\bibitem [{\citenamefont {Lin}\ \emph {et~al.}(2005)\citenamefont {Lin},
  \citenamefont {B{\"o}ker}, \citenamefont {Skaff}, \citenamefont {Cookson},
  \citenamefont {Dinsmore}, \citenamefont {Emrick},\ and\ \citenamefont
  {Russell}}]{lin2003nanoparticle}%
  \BibitemOpen
  \bibfield  {author} {\bibinfo {author} {\bibfnamefont {Y.}~\bibnamefont
  {Lin}}, \bibinfo {author} {\bibfnamefont {A.}~\bibnamefont {B{\"o}ker}},
  \bibinfo {author} {\bibfnamefont {H.}~\bibnamefont {Skaff}}, \bibinfo
  {author} {\bibfnamefont {D.}~\bibnamefont {Cookson}}, \bibinfo {author}
  {\bibfnamefont {A.~D.}\ \bibnamefont {Dinsmore}}, \bibinfo {author}
  {\bibfnamefont {T.}~\bibnamefont {Emrick}},\ and\ \bibinfo {author}
  {\bibfnamefont {T.~P.}\ \bibnamefont {Russell}},\ }\href
  {https://doi.org/10.1021/la048000q} {\bibfield  {journal} {\bibinfo
  {journal} {Langmuir}\ }\textbf {\bibinfo {volume} {21}},\ \bibinfo {pages}
  {191} (\bibinfo {year} {2005})}\BibitemShut {NoStop}%
\bibitem [{\citenamefont {Guzmán}\ \emph {et~al.}(2022)\citenamefont
  {Guzmán}, \citenamefont {Ortega},\ and\ \citenamefont {Rubio}}]{guzman2022}%
  \BibitemOpen
  \bibfield  {author} {\bibinfo {author} {\bibfnamefont {E.}~\bibnamefont
  {Guzmán}}, \bibinfo {author} {\bibfnamefont {F.}~\bibnamefont {Ortega}},\
  and\ \bibinfo {author} {\bibfnamefont {R.~G.}\ \bibnamefont {Rubio}},\ }\href
  {https://doi.org/10.1021/acs.langmuir.2c02038} {\bibfield  {journal}
  {\bibinfo  {journal} {Langmuir}\ }\textbf {\bibinfo {volume} {38}},\ \bibinfo
  {pages} {13313} (\bibinfo {year} {2022})}\BibitemShut {NoStop}%
\bibitem [{\citenamefont {Golestanian}\ \emph {et~al.}(1996)\citenamefont
  {Golestanian}, \citenamefont {Goulian},\ and\ \citenamefont
  {Kardar}}]{golestanian1999fluctuation}%
  \BibitemOpen
  \bibfield  {author} {\bibinfo {author} {\bibfnamefont {R.}~\bibnamefont
  {Golestanian}}, \bibinfo {author} {\bibfnamefont {M.}~\bibnamefont
  {Goulian}},\ and\ \bibinfo {author} {\bibfnamefont {M.}~\bibnamefont
  {Kardar}},\ }\href {https://doi.org/10.1103/PhysRevE.54.6725} {\bibfield
  {journal} {\bibinfo  {journal} {Phys. Rev. E}\ }\textbf {\bibinfo {volume}
  {54}},\ \bibinfo {pages} {6725} (\bibinfo {year} {1996})}\BibitemShut
  {NoStop}%
\bibitem [{\citenamefont {Noruzifar}\ and\ \citenamefont
  {Oettel}(2009)}]{nfr2009anis}%
  \BibitemOpen
  \bibfield  {author} {\bibinfo {author} {\bibfnamefont {E.}~\bibnamefont
  {Noruzifar}}\ and\ \bibinfo {author} {\bibfnamefont {M.}~\bibnamefont
  {Oettel}},\ }\href {https://doi.org/10.1103/PhysRevE.79.051401} {\bibfield
  {journal} {\bibinfo  {journal} {Phys. Rev. E}\ }\textbf {\bibinfo {volume}
  {79}},\ \bibinfo {pages} {051401} (\bibinfo {year} {2009})}\BibitemShut
  {NoStop}%
\bibitem [{\citenamefont {Noruzifar}\ \emph
  {et~al.}(2013{\natexlab{a}})\citenamefont {Noruzifar}, \citenamefont
  {Wagner},\ and\ \citenamefont {Zandi}}]{nfr2013three}%
  \BibitemOpen
  \bibfield  {author} {\bibinfo {author} {\bibfnamefont {E.}~\bibnamefont
  {Noruzifar}}, \bibinfo {author} {\bibfnamefont {J.}~\bibnamefont {Wagner}},\
  and\ \bibinfo {author} {\bibfnamefont {R.}~\bibnamefont {Zandi}},\ }\href
  {https://doi.org/10.1103/PhysRevE.87.020301} {\bibfield  {journal} {\bibinfo
  {journal} {Phys. Rev. E}\ }\textbf {\bibinfo {volume} {87}},\ \bibinfo
  {pages} {020301} (\bibinfo {year} {2013}{\natexlab{a}})}\BibitemShut
  {NoStop}%
\bibitem [{\citenamefont {Noruzifar}\ \emph
  {et~al.}(2013{\natexlab{b}})\citenamefont {Noruzifar}, \citenamefont
  {Wagner},\ and\ \citenamefont {Zandi}}]{nfr2013sct}%
  \BibitemOpen
  \bibfield  {author} {\bibinfo {author} {\bibfnamefont {E.}~\bibnamefont
  {Noruzifar}}, \bibinfo {author} {\bibfnamefont {J.}~\bibnamefont {Wagner}},\
  and\ \bibinfo {author} {\bibfnamefont {R.}~\bibnamefont {Zandi}},\ }\href
  {https://doi.org/10.1103/PhysRevE.88.042314} {\bibfield  {journal} {\bibinfo
  {journal} {Phys. Rev. E}\ }\textbf {\bibinfo {volume} {88}},\ \bibinfo
  {pages} {042314} (\bibinfo {year} {2013}{\natexlab{b}})}\BibitemShut
  {NoStop}%
\bibitem [{\citenamefont {Nellen}\ \emph {et~al.}(2009)\citenamefont {Nellen},
  \citenamefont {Helden},\ and\ \citenamefont
  {Bechinger}}]{nellen2009tunability}%
  \BibitemOpen
  \bibfield  {author} {\bibinfo {author} {\bibfnamefont {U.}~\bibnamefont
  {Nellen}}, \bibinfo {author} {\bibfnamefont {L.}~\bibnamefont {Helden}},\
  and\ \bibinfo {author} {\bibfnamefont {C.}~\bibnamefont {Bechinger}},\
  }\href@noop {} {\bibfield  {journal} {\bibinfo  {journal} {EPL (Europhysics
  Letters)}\ }\textbf {\bibinfo {volume} {88}},\ \bibinfo {pages} {26001}
  (\bibinfo {year} {2009})}\BibitemShut {NoStop}%
\bibitem [{\citenamefont {Gambassi}\ and\ \citenamefont
  {Dietrich}(2024)}]{gambassi2024critical}%
  \BibitemOpen
  \bibfield  {author} {\bibinfo {author} {\bibfnamefont {A.}~\bibnamefont
  {Gambassi}}\ and\ \bibinfo {author} {\bibfnamefont {S.}~\bibnamefont
  {Dietrich}},\ }\href@noop {} {\bibfield  {journal} {\bibinfo  {journal} {Soft
  Matter}\ }\textbf {\bibinfo {volume} {20}},\ \bibinfo {pages} {3212}
  (\bibinfo {year} {2024})}\BibitemShut {NoStop}%
\bibitem [{\citenamefont {Vasilyev}\ \emph {et~al.}(2011)\citenamefont
  {Vasilyev}, \citenamefont {Macio{\l}ek},\ and\ \citenamefont
  {Dietrich}}]{vasilyev2011critical}%
  \BibitemOpen
  \bibfield  {author} {\bibinfo {author} {\bibfnamefont {O.}~\bibnamefont
  {Vasilyev}}, \bibinfo {author} {\bibfnamefont {A.}~\bibnamefont
  {Macio{\l}ek}},\ and\ \bibinfo {author} {\bibfnamefont {S.}~\bibnamefont
  {Dietrich}},\ }\href@noop {} {\bibfield  {journal} {\bibinfo  {journal}
  {Physical Review E}\ }\textbf {\bibinfo {volume} {84}},\ \bibinfo {pages}
  {041605} (\bibinfo {year} {2011})}\BibitemShut {NoStop}%
\bibitem [{\citenamefont {Yolcu}\ and\ \citenamefont
  {Deserno}(2012)}]{deserno2012}%
  \BibitemOpen
  \bibfield  {author} {\bibinfo {author} {\bibfnamefont {C.}~\bibnamefont
  {Yolcu}}\ and\ \bibinfo {author} {\bibfnamefont {M.}~\bibnamefont
  {Deserno}},\ }\href {https://doi.org/10.1103/PhysRevE.86.031906} {\bibfield
  {journal} {\bibinfo  {journal} {Phys. Rev. E}\ }\textbf {\bibinfo {volume}
  {86}},\ \bibinfo {pages} {031906} (\bibinfo {year} {2012})}\BibitemShut
  {NoStop}%
\bibitem [{\citenamefont {Yolcu}\ \emph {et~al.}(2014)\citenamefont {Yolcu},
  \citenamefont {Haussman},\ and\ \citenamefont {Deserno}}]{yolcu2014}%
  \BibitemOpen
  \bibfield  {author} {\bibinfo {author} {\bibfnamefont {C.}~\bibnamefont
  {Yolcu}}, \bibinfo {author} {\bibfnamefont {R.~C.}\ \bibnamefont
  {Haussman}},\ and\ \bibinfo {author} {\bibfnamefont {M.}~\bibnamefont
  {Deserno}},\ }\href
  {https://doi.org/https://doi.org/10.1016/j.cis.2014.02.017} {\bibfield
  {journal} {\bibinfo  {journal} {Advances in Colloid and Interface Science}\
  }\textbf {\bibinfo {volume} {208}},\ \bibinfo {pages} {89} (\bibinfo {year}
  {2014})},\ \bibinfo {note} {special issue in honour of Wolfgang
  Helfrich}\BibitemShut {NoStop}%
\bibitem [{\citenamefont {Haussman}\ and\ \citenamefont
  {Deserno}(2014)}]{deserno2014}%
  \BibitemOpen
  \bibfield  {author} {\bibinfo {author} {\bibfnamefont {R.~C.}\ \bibnamefont
  {Haussman}}\ and\ \bibinfo {author} {\bibfnamefont {M.}~\bibnamefont
  {Deserno}},\ }\href {https://doi.org/10.1103/PhysRevE.89.062102} {\bibfield
  {journal} {\bibinfo  {journal} {Phys. Rev. E}\ }\textbf {\bibinfo {volume}
  {89}},\ \bibinfo {pages} {062102} (\bibinfo {year} {2014})}\BibitemShut
  {NoStop}%
\bibitem [{\citenamefont {Rahi}\ \emph {et~al.}(2009)\citenamefont {Rahi},
  \citenamefont {Emig}, \citenamefont {Graham}, \citenamefont {Jaffe},\ and\
  \citenamefont {Kardar}}]{emig2009}%
  \BibitemOpen
  \bibfield  {author} {\bibinfo {author} {\bibfnamefont {S.~J.}\ \bibnamefont
  {Rahi}}, \bibinfo {author} {\bibfnamefont {T.}~\bibnamefont {Emig}}, \bibinfo
  {author} {\bibfnamefont {N.}~\bibnamefont {Graham}}, \bibinfo {author}
  {\bibfnamefont {R.~L.}\ \bibnamefont {Jaffe}},\ and\ \bibinfo {author}
  {\bibfnamefont {M.}~\bibnamefont {Kardar}},\ }\href
  {https://doi.org/10.1103/PhysRevD.80.085021} {\bibfield  {journal} {\bibinfo
  {journal} {Phys. Rev. D}\ }\textbf {\bibinfo {volume} {80}},\ \bibinfo
  {pages} {085021} (\bibinfo {year} {2009})}\BibitemShut {NoStop}%
\bibitem [{\citenamefont {Rodriguez-Lopez}\ \emph {et~al.}(2015)\citenamefont
  {Rodriguez-Lopez}, \citenamefont {Emig}, \citenamefont {Noruzifar},\ and\
  \citenamefont {Zandi}}]{emig2015}%
  \BibitemOpen
  \bibfield  {author} {\bibinfo {author} {\bibfnamefont {P.}~\bibnamefont
  {Rodriguez-Lopez}}, \bibinfo {author} {\bibfnamefont {T.}~\bibnamefont
  {Emig}}, \bibinfo {author} {\bibfnamefont {E.}~\bibnamefont {Noruzifar}},\
  and\ \bibinfo {author} {\bibfnamefont {R.}~\bibnamefont {Zandi}},\ }\href
  {https://doi.org/10.1103/PhysRevA.91.012516} {\bibfield  {journal} {\bibinfo
  {journal} {Phys. Rev. A}\ }\textbf {\bibinfo {volume} {91}},\ \bibinfo
  {pages} {012516} (\bibinfo {year} {2015})}\BibitemShut {NoStop}%
\bibitem [{\citenamefont {Noruzifar}\ \emph
  {et~al.}(2013{\natexlab{c}})\citenamefont {Noruzifar}, \citenamefont
  {Rodriguez-Lopez}, \citenamefont {Emig},\ and\ \citenamefont
  {Zandi}}]{nfr2013}%
  \BibitemOpen
  \bibfield  {author} {\bibinfo {author} {\bibfnamefont {E.}~\bibnamefont
  {Noruzifar}}, \bibinfo {author} {\bibfnamefont {P.}~\bibnamefont
  {Rodriguez-Lopez}}, \bibinfo {author} {\bibfnamefont {T.}~\bibnamefont
  {Emig}},\ and\ \bibinfo {author} {\bibfnamefont {R.}~\bibnamefont {Zandi}},\
  }\href {https://doi.org/10.1103/PhysRevA.87.042504} {\bibfield  {journal}
  {\bibinfo  {journal} {Phys. Rev. A}\ }\textbf {\bibinfo {volume} {87}},\
  \bibinfo {pages} {042504} (\bibinfo {year} {2013}{\natexlab{c}})}\BibitemShut
  {NoStop}%
\bibitem [{\citenamefont {Noruzifar}\ \emph {et~al.}(2012)\citenamefont
  {Noruzifar}, \citenamefont {Emig}, \citenamefont {Mohideen},\ and\
  \citenamefont {Zandi}}]{nfr2012}%
  \BibitemOpen
  \bibfield  {author} {\bibinfo {author} {\bibfnamefont {E.}~\bibnamefont
  {Noruzifar}}, \bibinfo {author} {\bibfnamefont {T.}~\bibnamefont {Emig}},
  \bibinfo {author} {\bibfnamefont {U.}~\bibnamefont {Mohideen}},\ and\
  \bibinfo {author} {\bibfnamefont {R.}~\bibnamefont {Zandi}},\ }\href
  {https://doi.org/10.1103/PhysRevB.86.115449} {\bibfield  {journal} {\bibinfo
  {journal} {Phys. Rev. B}\ }\textbf {\bibinfo {volume} {86}},\ \bibinfo
  {pages} {115449} (\bibinfo {year} {2012})}\BibitemShut {NoStop}%
\bibitem [{\citenamefont {Noruzifar}\ \emph {et~al.}(2011)\citenamefont
  {Noruzifar}, \citenamefont {Emig},\ and\ \citenamefont {Zandi}}]{nfr2011}%
  \BibitemOpen
  \bibfield  {author} {\bibinfo {author} {\bibfnamefont {E.}~\bibnamefont
  {Noruzifar}}, \bibinfo {author} {\bibfnamefont {T.}~\bibnamefont {Emig}},\
  and\ \bibinfo {author} {\bibfnamefont {R.}~\bibnamefont {Zandi}},\ }\href
  {https://doi.org/10.1103/PhysRevA.84.042501} {\bibfield  {journal} {\bibinfo
  {journal} {Phys. Rev. A}\ }\textbf {\bibinfo {volume} {84}},\ \bibinfo
  {pages} {042501} (\bibinfo {year} {2011})}\BibitemShut {NoStop}%
\bibitem [{\citenamefont {Liu}\ \emph {et~al.}(2018)\citenamefont {Liu},
  \citenamefont {Sharifi-Mood},\ and\ \citenamefont {Stebe}}]{liu2018}%
  \BibitemOpen
  \bibfield  {author} {\bibinfo {author} {\bibfnamefont {I.~B.}\ \bibnamefont
  {Liu}}, \bibinfo {author} {\bibfnamefont {N.}~\bibnamefont {Sharifi-Mood}},\
  and\ \bibinfo {author} {\bibfnamefont {K.~J.}\ \bibnamefont {Stebe}},\ }\href
  {https://doi.org/https://doi.org/10.1146/annurev-conmatphys-031016-025514}
  {\bibfield  {journal} {\bibinfo  {journal} {Annual Review of Condensed Matter
  Physics}\ }\textbf {\bibinfo {volume} {9}},\ \bibinfo {pages} {283} (\bibinfo
  {year} {2018})}\BibitemShut {NoStop}%
\bibitem [{\citenamefont {Zhou}\ \emph {et~al.}(2020)\citenamefont {Zhou},
  \citenamefont {Liu}, \citenamefont {Zhang}, \citenamefont {Miao},
  \citenamefont {Luo},\ and\ \citenamefont {Jing}}]{zhou_2020}%
  \BibitemOpen
  \bibfield  {author} {\bibinfo {author} {\bibfnamefont {M.}~\bibnamefont
  {Zhou}}, \bibinfo {author} {\bibfnamefont {Y.}~\bibnamefont {Liu}}, \bibinfo
  {author} {\bibfnamefont {P.}~\bibnamefont {Zhang}}, \bibinfo {author}
  {\bibfnamefont {Y.}~\bibnamefont {Miao}}, \bibinfo {author} {\bibfnamefont
  {H.}~\bibnamefont {Luo}},\ and\ \bibinfo {author} {\bibfnamefont
  {G.}~\bibnamefont {Jing}},\ }\href {https://doi.org/10.1088/1367-2630/ab7f90}
  {\bibfield  {journal} {\bibinfo  {journal} {New Journal of Physics}\ }\textbf
  {\bibinfo {volume} {22}},\ \bibinfo {pages} {053005} (\bibinfo {year}
  {2020})}\BibitemShut {NoStop}%
\bibitem [{\citenamefont {Villanueva-Valencia}\ \emph
  {et~al.}(2021)\citenamefont {Villanueva-Valencia}, \citenamefont {Leferink},
  \citenamefont {Veen}, \citenamefont {Huerta}, \citenamefont {Dullens},\ and\
  \citenamefont {de~las Heras}}]{villanueva2021concentration}%
  \BibitemOpen
  \bibfield  {author} {\bibinfo {author} {\bibfnamefont {J.~R.}\ \bibnamefont
  {Villanueva-Valencia}}, \bibinfo {author} {\bibfnamefont {V.~G.}\
  \bibnamefont {Leferink}}, \bibinfo {author} {\bibfnamefont {S.}~\bibnamefont
  {Veen}}, \bibinfo {author} {\bibfnamefont {A.}~\bibnamefont {Huerta}},
  \bibinfo {author} {\bibfnamefont {R.~P.~A.}\ \bibnamefont {Dullens}},\ and\
  \bibinfo {author} {\bibfnamefont {D.}~\bibnamefont {de~las Heras}},\
  }\href@noop {} {\bibfield  {journal} {\bibinfo  {journal} {Physical Chemistry
  Chemical Physics}\ }\textbf {\bibinfo {volume} {23}},\ \bibinfo {pages}
  {4404} (\bibinfo {year} {2021})}\BibitemShut {NoStop}%
\bibitem [{\citenamefont {Mousavi}\ and\ \citenamefont
  {Noruzifar}(2025)}]{zenododata}%
  \BibitemOpen
  \bibfield  {author} {\bibinfo {author} {\bibfnamefont {S.~E.}\ \bibnamefont
  {Mousavi}}\ and\ \bibinfo {author} {\bibfnamefont {E.}~\bibnamefont
  {Noruzifar}},\ }\href {https://doi.org/10.5281/zenodo.17257298}
  {10.5281/zenodo.17257298} (\bibinfo {year} {2025})\BibitemShut {NoStop}%
\end{thebibliography}%
	
\end{document}